# Strain Induced Robust Skyrmion lattice at Room Temperature in van der Waals Ferromagnet

*Xinyi Zhou [#], Iftikhar Ahmed Malik [#], Ruihuan Duan[#], Hanqing Shi, Chen Liu, Yan Luo, Yue Sun, Ruixi Chen, Yilin Liu, Shian Xia, Vanessa Li Zhang, Sheng Liu[\*], Chao Zhu, Xixiang Zhang, Yi Du, Zheng Liu[\*], Ting Yu[\*]*

\# These authors contributed equally to this work.

Xinyi Zhou, Iftikhar Ahmed Malik, Yue Sun, Ruixi Chen, Yilin Liu, Shian Xia, Vanessa Li Zhang, Sheng Liu, Ting Yu

School of Physics and Technology

Wuhan University

Wuchang District, Hubei 430072, China.

E-mail: yu.ting@whu.edu.cn; liu.sheng@whu.edu.cn

Ting Yu

Wuhan Institute of Quantum Technology

Wuhan 430206, China.

Ting Yu

Key Laboratory of Artificial Micro- and Nano- structures of Ministry of Education

Wuhan University

Wuhan 430072, China.

Ruihuan Duan, Zheng Liu

School of Materials Science and Engineering

Nanyang Technological University

Singapore 639798, Singapore.

E-mail: z.liu@ntu.edu.sg

Hanqing Shi, Yi Du

School of Physics

Beihang University

Haidian District, Beijing 100191, China.

Chen Liu, Xixiang Zhang

Physical Science and Engineering Division (PSE)

King Abdullah University of Science and Technology (KAUST)

Thuwal, 23955-6900 Saudi Arabia.

Yan Luo, Chao Zhu

SEU-FEI Nano-Pico Center, Key Lab of MEMS of Ministry of Education

School of Integrated Circuits

Southeast University

Nanjing 210096, China.

**Abstract**

Manipulating topological magnetic orders of two-dimensional (2D) magnets by strain, once achieved, offers enormous potential for future low-power flexible spintronic applications. In this work, by placing $Fe_3GaTe_2$ (FGaT), a room-temperature 2D ferromagnet, on flexible substrate, we demonstrate a field-free and robust formation of skyrmion lattice induced by strain. By applying a minimal strain of ~0.80% to pre-annealed FGaT flakes, the Magnetic Force Microscopy (MFM) tip directly triggers the transition from maze-like domains to an ordered skyrmion lattice while scanning the sample surface. The skyrmion lattice is rather stable against extensive cyclic mechanical testing (stretching, bending, and twisting over 2000 cycles each). It also exhibited stability across a wide range of magnetic fields (~2.9 kOe) and temperatures (~ 323 K), as well as long-term retention stability, highlighting its robustness and field free stabilization. The strain effect reduces the lattice symmetry and enhances the Dzyaloshinskii-Moriya interaction (DMI) of FGaT, thus stabilizing the skyrmion lattice.

Our findings highlight the potential of FGaT for integrating magnetic skyrmions into future low-power-consumption flexible spintronics devices.

## 1. Introduction

Topologically stabilized magnetic spin structures at the nanoscale, including domain walls[1], vortices[2] and skyrmions[3], have recently received much attention where structure and dynamics of sub-micrometer magnetic domains are the main factors determining the physical properties and applications. Among them, magnetic skyrmions—topologically non-trivial swirling spin configurations arising from the competing magnetic interactions—have captured significant attention for their potential in low-power spintronic memory and computational devices[4]. To serve as information carriers, skyrmions require reliable and controllable creation methods ("writing") triggered by external stimuli. To date, various methods for manipulating skyrmions have been investigated, including the application of electric current[5], magnetic field[6], and laser pulses[7]. Recent advances in flexible devices have highlighted strain engineering as an innovative and energy-efficient alternative for generation and control of skyrmions, enabling novel applications in reconfigurable non-volatile memory and logic devices[8]. However, so far, due to the difficulty in operating and controlling the methods of applying sufficient strain to overcome the energy barrier between skyrmions and other magnetic states, the manipulation of skyrmion formation via strain without external magnetic field, especially the generation of robust skyrmion lattice at room temperature, has not been fully explored[9].

It is well known that strain plays a significant role in regulating the DMI[10], which generally arises from the breaking of inversion symmetry in non-centrosymmetric materials. Some studies have demonstrated that DMI can also be introduced in centrosymmetric materials through strain engineering[11]. Epitaxial growth on substrates with controlled lattice constant mismatch has long been employed to induce built-in strain, but its effectiveness is significantly reduced for 2D materials due to their inherently weak substrate interactions[12]. A key challenge in strain engineering of 2D materials through ferroelectric (FE) substrate is weak interfacial adhesion, resulting in inefficient strain transfer[13]. Additionally, FE substrates may introduce complexities such as polarization-induced charges at the interface, which can affect the skyrmion formation[14]. Recently, surface acoustic waves have emerged as a promising tool for manipulating skyrmions generation and dynamics, but their implementation relying on piezoelectric substrates introduces not only controllable strain effects but also significant unintended thermal perturbations and persistent charge-accumulation phenomena[15]. Furthermore, most ferromagnetic materials studied for skyrmions either involve multilayer thin films with complex interfacial quality, film construction, and precise thickness control hurdle[9, 16], or a limitation of a low Curie temperature ($T_c$) below room temperature, significantly hindering their practical applications[10a, 10c]. Therefore, achieving field-free, ordered, and room-temperature strain-induced skyrmion lattices in 2D van der Waal (vdW) ferromagnets remains a critical challenge in advancing practical spintronic applications.

In this study, by applying tensile strain to pre-annealed FGaT on polyethylene terephthalate (PET) substrate, we demonstrate the successful formation of robust skyrmion lattice with uniform skyrmions size at room temperature. This skyrmion lattice was directly observed using an in-situ strain MFM, through a series of controlled experiments with varying pre-annealing temperatures and strain values. Pre-annealing at 323 K enables the formation of an ordered skyrmion lattice with only 0.80% tensile strain. This magnetic skyrmion lattice remains stable after undergoing numerous cyclic mechanical tests, including stretching, bending, and twisting over 2000 cycles each. Meanwhile, it exhibits stability over a wide range of magnetic fields and temperatures, and maintains long-term stability, highlighting the robustness of the lattice and its ability to stabilize without the need for external fields. In contrast, the skyrmions formed by field cooling (FC) exhibited a pre-disordered distribution. Notably, pre-annealed FGaT at lower or higher temperatures did not show such robust skyrmion lattice, highlighting the existence of an optimal window of pre-annealing temperature. For these optimized pre-annealed FGaT samples, the high-angle annular dark-field scanning transmission electron microscopy (HAADF-STEM) and Raman spectroscopy measurements confirm the formation of $FeTe_2$ layers on their surface, which disrupted the lattice symmetry and enhanced the DMI, thus fascinating the strain induced generation of skyrmion lattice.

## 2. Results and Discussion

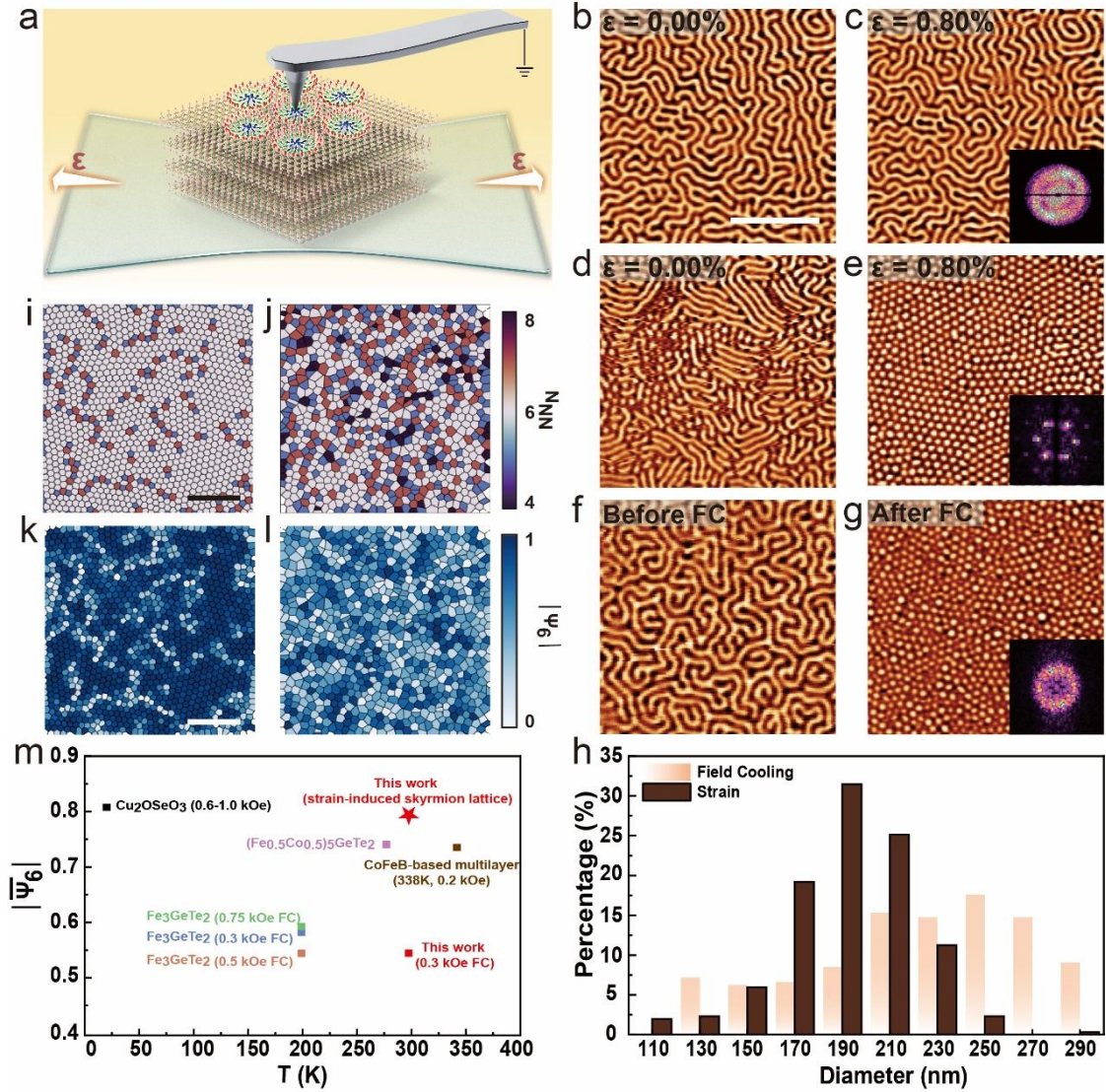

**Figure 1 | Strain-induced skyrmion lattice. a,** Schematic diagram of in-situ uniaxial strain-induced skyrmion lattice formation at room-temperature on flexible PET substrate. **b, c,** MFM images of non-annealed sample before (**b**) and after (**c**) applying 0.80% strain. The bottom right inset in **c** shows the corresponding fast Fourier transform (FFT) map, which exhibits no symmetry. **d, e,** MFM images of 323 K pre-annealed sample before (**d**) and after (**e**) applying 0.80% strain. The bottom right inset in **e** shows the corresponding FFT map, which exhibits a six-fold symmetry, indicating the ordering of skyrmion lattice. **f, g,** MFM images before (**f**) and after FC (**g**), respectively. **h**, Statistical comparison of percentage of skyrmions of different sizes generated by strain method (**e**) and FC method (**g**). **i, j,** Voronoi tessellation highlights the nearest neighbors ($N_{NN}$) of each skyrmion for strain-induced skyrmion lattice phase (**i**) and disordered skyrmion phase by FC (**j**), respectively. **k, l,** the bond orientational parameter ($|\Psi_6|$) of strain-induced skyrmion lattice phase (**k**) and disordered skyrmion phase by FC (**l**),

respectively. **m,** Comparison of average $|\Psi_6|$ in various materials[17] with this study. Scale bar, 2 $\mu m$.

## 2.1. Strain-Induced Skyrmion Lattice

One evident effect of applying strain on FGaT is the transition from maze-like domains to an ordered skyrmion lattice, compared to disordered skyrmions generated during typical FC process, as observed through in-situ mechanical strain MFM on a custom-designed strain stage (schematically depicted in **Figure 1**a) at room temperature. Figure 1 illustrates a comprehensive study of skyrmion formation in FGaT in response to strain and FC. Thermally pre-annealed FGaT samples at different temperatures, were mechanically exfoliated onto flexible PET substrates as detailed in Figure S1. In the non-annealed sample (Figures 1b, c), maze-like domains remained unchanged before and after applying strain.

A notable transformation from maze-like domains (Figure 1d) to a well-ordered skyrmion lattice (Figure 1e) was observed in the 323 K pre-annealed sample. Prior to the application of strain, certain skyrmions are also observed in strain free state (Figure 1d) which led us to hypothesize that minor strains incurred during the transfer of the flakes from polydimethylsiloxane (PDMS) to PET are responsible for the generation of early skyrmions (more details are shown in Figure S16). Upon increasing the strain to 0.80%, highly ordered skyrmion lattice is formed, as shown in Figure 1e. In comparison, field-free disordered skyrmion phase (Figure 1g) is observed after 0.3 kOe FC process, where distinct contrast corresponds to different sized skyrmions. Pristine FGaT sample with maze-like domains phase was first heated to temperature higher than $T_C$ of FGaT which were then naturally cooled down to room temperature under 0.3 kOe for FC process. Figure 1e, g, were subjected to fast Fourier transform (FFT) analysis (Inset of Figure 1e, g) which revealed a six-fold coordination arrangement for strain-induced skyrmion lattice (Figure 1e), further confirming a short-range ordering. However, FFT analysis of skyrmion phase generated by FC (inset Figure 1g) exhibits an oval shape indicating relatively low symmetry. Figure 1h presents a histogram comparing the size distributions of skyrmions induced by strain and FC. The data revealed a marked uniformity in skyrmion sizes produced through the application of strain, with the majority clustering around 200 nm in diameter. In contrast, the size distribution of skyrmions generated by FC is notably broader, spanning a wide range of diameters from 110 nm to nearly 290 nm. A detailed statistical analysis was performed to analyze the ordered/disordered skyrmion phases. Utilizing an algorithm to identify skyrmions in MFM images of observed phases[17], Voronoi tessellation was applied to highlight the nearest neighbors ($N_{NN}$) of each skyrmion as shown in Figure 1i, j, corresponding to the strain-induced ordered skyrmion lattice phase and the disordered skyrmion phase obtained through FC, respectively. In the strain-induced skyrmion lattice phase, most lattice sites exhibit the 6-fold coordination, only disrupted by pairs of correlated dislocations (neighboring pairs of 5- and 7-fold sites) with opposite Burgers' vectors[18]. Voronoi tessellation of strain-induced skyrmion lattice shows a few 5- and 7- fold defect sites indicating a predominantly solid (ordered) phase[17a]. In contrast, the skyrmion phase produced by the FC process shows significantly fewer 6-fold-coordinated sites, indicative of a liquid (disordered) phase. The bond orientational parameter ($|\Psi_6|$) further quantifies these phases, where values of 1(or near 0) represent a solid(liquid) phase[17b]. The quasi-global measure of short-range lattice order was obtained by averaging the $|\Psi_6|$ values of all skyrmions in the image. Figure 1k demonstrates that the strain-induced skyrmion lattice has an average $|\overline{\Psi_6}| = 0.74$, suggesting a solid phase, whereas Figure 1l shows that the FC-induced skyrmion phase has a lower value of 0.54, indicating a denser and more disordered liquid phase. Notably, skyrmions observed by FC may exhibit complex features inconsistent with 2D Néel-type skyrmions as suggested theoretically in $Fe_3GeTe_2$[19] and frustrated systems[20]. This inconsistency can be resolved if skyrmions observed by FC are considered to be bent/twisted 3D spin textures giving rise to random repulsive interactions and disordered skyrmions phase[21]. Compared with isolated skyrmions, the skyrmion lattice is a periodic arrangement composed of multiple skyrmions. Due to the coupling effect between them, the stability of a single skyrmion is enhanced, making the skyrmion lattice highly resistant to external interference and thus having a greater overall structural stability. The high periodicity of skyrmion lattice enables large-scale spin alignment and information storage in materials. Therefore, by controlling the arrangement and interaction of the skyrmion lattice, more efficient storage and computing applications can be achieved[22]. As summarized in Figure 1m, several studies have employed the average $|\Psi_6|$ value as a metric to quantify the skyrmion lattice ordering in various materials. The strain-induced skyrmion lattice analyzed in this study exhibits an exceptionally high degree of orderliness at room temperature, emphasizing its significant potential for both research and practical applications[17].

For a magnetic system hosting topological spin configurations, the Hall resistance ($R_{xy}^{Hall}$) consists of the normal, anomalous, and topological Hall resistances, where topological Hall resistance ($R_{xy}^T$) is associated with a noncollinear spin texture that comprises the skyrmions[23]. To investigate the potential existence of $R_{xy}^T$ in FGaT we fabricated a Hall device with a sample thickness of 300 nm and conducted measurements of $R_{xy}$ with the external magnetic field (*H*) applied perpendicularly to the sample surface at 300 K (see details in Supporting Note 1). $R_{xy}$(*H*) curve shows a sheared out-of-plane hysteresis loop with two distinct slopes. The result suggests the presence of non-linear spin textures. MFM measurements with varying magnetic field (as shown in Figure S2) can precisely indicate the occurrence of skyrmions during the process of changing magnetic field. Combined with the *M*(*H*) curve (Figure S3b) obtained from reflective magnetic circular dichroism (RMCD) testing, the $R_{xy}^T$ is extracted through data processing, as displayed in Figure S3c. There is a discrepancy at the low-field region, indicating the presence of a pronounced $R_{xy}^T$ component in $R_{xy}$, which proves the existence of topological spin configurations in FGaT samples.

## 2.2. Strain-Pre-annealing Temperature Phase Diagram

To explore how magnetic domains evolve with pre-annealing temperature and strain, a series of MFM measurements were conducted on FGaT samples of uniform thickness (**Figure 2**a). In the non-annealed sample, even with 0.80% strain applied, the maze-like magnetic domains remained unchanged. However, in samples pre-annealed at 313 K, a transformation from maze-like domains to skyrmion phases is observed at 0.40% strain and a fully ordered skyrmion lattice is formed at 0.80% strain. Notably, sample annealed at 323 K showed a strong response to minor mechanical disturbances, consistently forming a well-ordered skyrmion lattice. Furthermore, as shown in Figure S6, it has been experimentally demonstrated that the formation of skyrmions exhibits a proportional increase in strain sensitivity with extended thermal processing time. Whereas, at higher pre-annealing temperatures (333 K and 343 K), there exists a more ordered striped pattern with few skyrmions and higher-ordered nonlinear spin textures, such as skyrmionium. The MFM images in these high-temperature pre-annealed samples also displayed complex domain structures, with regions of different contrast, that is a stronger contrast in ferromagnetic domains and a weaker contrast in reduced ferromagnetism regions. More MFM images corresponding to the phase diagram shown in Figure 2a are provided in Figure S4-8.

To calibrate strain amplitude quantitively, we designed a graphene-FGaT-graphene sandwich structure. The strain is obtained and analyzed on the top and bottom surfaces of FGaT by examining the 2D peak shift in monolayer graphene, as shown in Figure S9. Uniaxial strain shifts the frequency of the 2D band in monolayer graphene due to changes in lattice vibrations, providing a method for strain quantification[24] The observed redshift of the 2D peak for the bottom graphene, in contrast to the absence of any shift in the top graphene, suggests the presence of inhomogeneous strain within the FGaT layer (Figure S9b). The inset in Figure 2c shows the Raman spectra of the 2D band of bottom monolayer graphene under varying strains, with a redshift observed as strain increases. Figure 2c quantifies the linear relationship between the 2D band frequency and strain, enabling precise strain measurement in graphene (detailed in Figure S9-11).

The phase diagram in Figure 2b outlines the specific temperature-strain window essential for skyrmion formation in FGaT. It identifies the critical strain and pre-annealing temperature conditions required to stabilize skyrmions. It is observed that the magnetic domains on either side of the temperature window shows significant differences in shape, with maze-like domains at 303 K and more ordered striped domains at higher temperatures (333 K and 343 K). The distinct observation in magnetic domains suggests the change in magnetic energies (Supporting Note 2) before and after pre-annealing as the domains pattern arises from the competition between short-range magnetic interactions and long-range dipolar interactions.

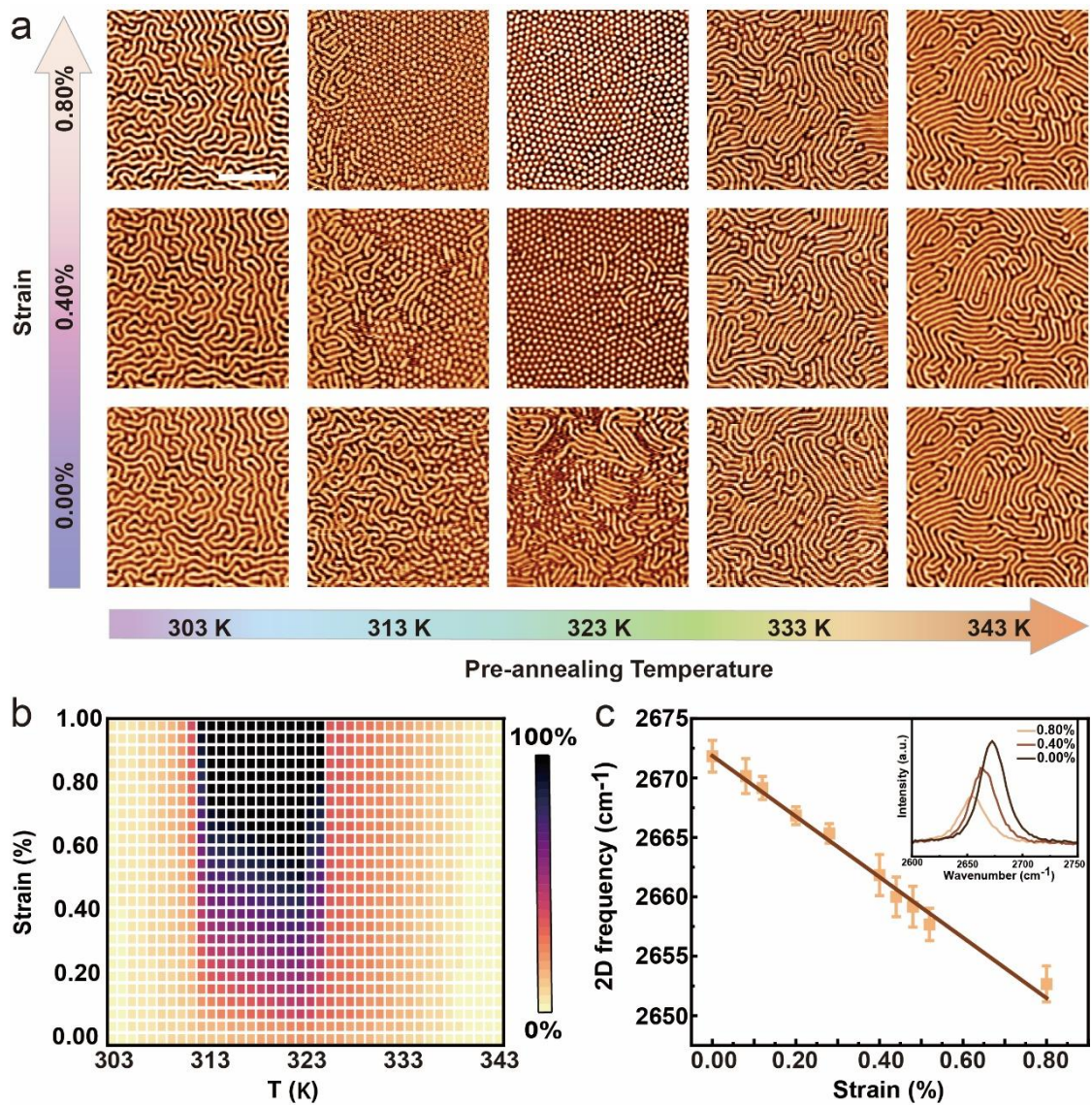

**Figure 2 | Skyrmion lattice evolution under strain and pre-annealing temperatures phase diagram. a**, Evolution of skyrmion lattice in 280 nm thick FGaT sample at different values of strain and pre-annealing temperature. Scale bar, 2 $\mu m$. **b**, Magnetic phase diagram deduced from experimental data acquired by MFM. The color indicates the percentage of the scanned image area that has transformed into skyrmions. **c**, Calibration of uniaxial strain on PET using 2D peak positions of monolayer graphene. Inset **(c)** is Raman spectra of monolayer graphene on PET substrate at different strain values.

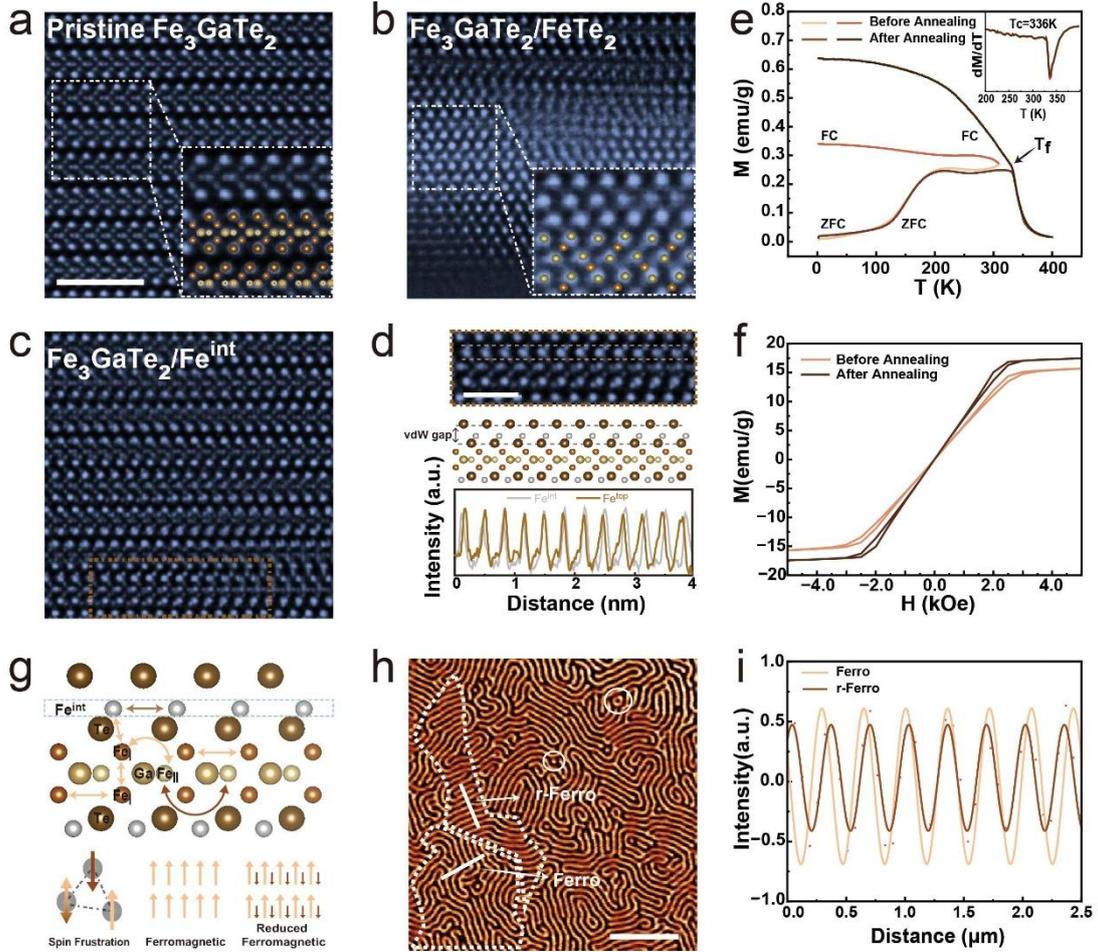

**Figure 3 | Structural transformations and underlying mechanisms at various pre-annealing temperatures. a-d,** STEM images of FGaT before **(a)** and after pre-annealing **(b)** viewed along [001] orientation. Inset shows magnified view of the atomic structure of FGaT (in Figure 3a) and FeTe$_2$ (Figure 3b) overlapped with the schematic. STEM image highlighting the intercalated Fe atoms (Fe$^{int}$) within the vdW gaps of FGaT after high-temperature pre-annealing **(c)**. Zoom-in STEM image of the brown dotted rectangle from **(c)**, showing Fe$^{int}$ atom in vdW gaps of FGaT, along with schematic of atomic arrangements and corresponding intensity line profiles of Fe$^{int}$ and Fe$^{top}$ atoms, represented by gray and brown curves, respectively **(d)**. **e,** M-T curves measured after zero field cooling (ZFC) and FC before and after pre-annealing. **f,** M-H curves measured at room temperature before and after pre-annealing. **g,** Schematic of pre-annealed bilayer FGaT illustrating local Fe$^{int}$ in vdW gap and considered exchange paths of the interatomic forces with orange representing ferromagnetic interactions and brown representing antiferromagnetic interactions. The geometric spin frustration within a triangular sublattice is also shown, where the mismatched interactions prevent the system from achieving global energy minimization through simple spin alignment, resulting in a spin-frustrated state. **h,** MFM image of magnetic domains in 343 K pre-annealed FGaT flake, showing two different phases with strong (ferromagnetic phase) and weak (reduced ferromagnetic phase) contrast, with phase boundaries marked by dotted loops. **i,** The line profile

of the two-phase regions (Ferro and r-Ferro) from **h**. Scale bar in **a** and **d** is 2 nm and 2 $\mu m$ in **h**.

### 2.3. Effects of Pre-annealing at Various Temperatures

To elucidate the origin of the pre-annealing temperature window and investigate pre-annealing effects on FGaT crystalline structures, the high-angle annular HAADF-STEM was performed on cross-sectional samples before and after pre-annealing (**Figure 3**a-d). The observations reveal that the pristine FGaT exhibits a vdWs stacking structure, with its crystal structure identified in the P6$_3$/mmc space group[25] (Figure 3a and Figure S12). In each FGaT unit cell, Fe$_3$Ga slab is sandwiched by two Te layers. Within the sublayer, Fe$_I$ (at the top and bottom) and Fe$_{II}$ (at the center) atoms occupy two unique Wyckoff positions.

The STEM image in Figure 3b shows the formation of non-uniform FeTe$_2$, after low temperature annealing (313 K and 323 K), which locally disrupts the symmetry of FGaT. Inset shows a magnified view of the atomic structure of FeTe$_2$ overlaid with the schematic. Raman spectroscopy further confirmed precipitation as discussed in Figure S13. A broad peak around 150 cm$^{-1}$ is observed in the pristine FGaT sample, representing the lattice vibrational modes (A$_{1g}$+E$_{2g}$) of FGaT. As the pre-annealing temperature increases, the broad peak gradually changes. At high temperatures, two sharp peaks of FeTe$_2$ at 121 cm$^{-1}$ (A$_g$) and 140 cm$^{-1}$ (B$_{1g}$) appear in the Raman spectrum, indicating changes in the FGaT crystal structure. Raman mapping of A$_g$ (Figure S13c) and B$_{1g}$ (Figure S13d) mode intensity indicates the presence of FeTe$_2$ across the entire FGaT flake.

Additionally, accumulation of Fe$^{int}$ within the vdW gaps of FGaT after high-temperature pre-annealing (403 K) is observed in STEM image (Figure 3c). Figure 3d shows a zoom-in STEM image of the brown-dotted rectangle from Figure 3c, along with the schematic of atomic arrangements and the corresponding intensity line profiles of Fe$^{int}$ and Fe$_I$ atoms, represented by gray and brown curves, respectively. The random distribution of Fe$^{int}$ significantly affects the macroscopic magnetic properties of FGaT, as indicated by macroscopic magnetization measurements (Figure 3e and f). The disordered spins cause a bifurcation between the zero-field cooled (ZFC) and FC temperature dependent magnetization (M-T) curves (Figure 3c) at a spin-freezing temperature ($T_f$). An additional kink-like feature at $T_f$ in M-T curve of annealed sample likely results from antiferromagnetically coupled disordered spins. Enhanced magnetization observed in isothermal magnetization curves (M-H) measured after pre-annealing (Figure 3f) also results from disordered spins and strong ferromagnetic coupling from $Fe^{int}$ [26].

Magnetic coupling between iron atoms in FGaT and the nature of disordered spins has been understood by density functional theory[26-27]. Strong ferromagnetic coupling is preferred among nearest neighbor iron atoms within the $Fe_3Ga$ sublayer and third-nearest neighbor interaction ($F_{II}$-$F_{II}$) favors antiferromagnetic coupling. Furthermore, interactions between $Fe^{int}$ atoms favor strong antiferromagnetic coupling. These antiferromagnetically coupled iron atoms are arranged in triangular lattices, inducing spin frustration as shown in Figure 3g. The effects of structural changes and geometric frustration on the magnetic domains is evident in the MFM image (Figure 3h) obtained at room temperature after high-temperature pre-annealing. Topological structures such as skyrmions and skyrmioniums, along with noticeable non-uniform MFM contrast is observed revealing complex pattern consisting of two distinct regions (stronger and weaker MFM contrast). When an antiferromagnetic phase and a ferromagnetic phase coexist, the magnetic moments of the ferromagnetic phase may experience antiferromagnetic interactions in specific regions or under certain conditions. This interaction disrupts the highly ordered and uniform alignment characteristic of a pure ferromagnetic phase, leading to weakened magnetic moments or partial disorder in the ferromagnetic regions. Such behavior can be analogously described as " reduced ferromagnetism". The weak contrast regions (Figure 3h) are attributed to the coexistence of interlayer antiferromagnetism with a ferromagnetic background (labeled as r-Ferro), while the strong contrast regions (Figure 3h) correspond to the predominant ferromagnetic domains (labeled as Ferro). The line profiles of the two-phase regions (Ferro and r-Ferro) are shown in Figure 3i clearly distinguishing the contrast from Ferro and r-Ferro phases.

**2.4. Roles of Strain and Pre-annealing in Skyrmion Lattice Formation**

The lattice of FGaT was generally reported to belong to centrosymmetric $P6_3/mmc$ space group, thus in principle it should not exhibit a global DMI. In that case, maze-like domains remained unchanged before and after applying strain, in the non-annealed sample (Figure 1b, c; Figure 2a), dominated by exchange energy and dipolar interactions[28].

Recent studies suggest that due to the Fe vacancies[28], the FGaT lattice might be in a non-centrosymmetric $P3m1$ space group[7c, 29]. The existence of surface oxidized layer[29b] or inhomogeneous $Fe^{int}$ [30] might lead to a global DMI, resulting Néel-type character of domain wall and/or skyrmions. However, oxidation was excluded as a cause of DMI in this study. As illustrated in Figures S14, when FGaT sample was partially protected with thin hexagonal boron nitride (hBN) within a high-purity nitrogen-filled glove box and subjected to the conventional FC method, skyrmions were observed in both non-covered FGaT and the hBN/FGaT heterostructure. This suggests that the DMI originates from intrinsic symmetry breaking rather than interface oxidation. Furthermore, all strain-induced skyrmion lattice experiments, including sample exfoliation, pre-annealing, strain application, and MFM testing, were conducted within the glove box. The formation of regular skyrmion lattice was consistently observed (Figures S15). It is worth noting that the pre-annealing created precipitation of $FeTe_2$ near the surface (Figure 3b) and $Fe^{int}$ within the vdW gaps (Figure 3c) of FGaT. This $FeTe_2$/FGaT heterostructure has been proposed to break inversion symmetry, potentially enhancing the DMI through interfacial asymmetry[31]. These results indicate that the skyrmion lattice in FGaT arises from inhomogeneous strain-enhanced DMI rather than oxygen interface effects. In the low temperature (313 K and 323 K) pre-annealed cases, pre-annealing led to a transition from the maze-like domain phase to a stripe phase (Figure 1d), minimizing domain wall energy, reducing dipolar interactions. These magnetic energy optimizations made FGaT more sensitive to strain and external magnetic fields, leading to a DMI-dominated state.

As explained earlier, strain engineering has emerged as a powerful tool for modulating the DMI and enhancing chiral spintronics in magnetic materials. It has been shown that DMI can be achieved in a centrosymmetric material by inhomogeneous strain[32]. Thus, the application of inhomogeneous strain can further break the symmetry, reduce the energy barrier between the stripe domain and the skyrmion lattice state, enabling the local stray magnetic field of the MFM tip to facilitate this transition. Upon the application of inhomogeneous strain on 323 K pre-annealed FGaT and MFM tip stray field, the system stabilized into a robust, field-free skyrmion lattice. However, without applying strain, only pre-annealing the sample at appropriate temperatures, multiple scans with the MFM magnetic tip failed to form skyrmion lattice (Figure S16). This is also evidenced by Figure S17. When in-situ strain was applied via blue tape during sample preparation, the formation of the skyrmion lattice was observed. This also accounts for the formation of skyrmions in some initial states (0.00% strain), caused by uncontrolled strain during sample preparation. The possible mechanism that led to observation of strain induced skyrmion lattice is briefly illustrated in the magnetic field-strain energy landscape diagrams (Figure S18). In non-annealed FGaT, the maze-like domain phase ($S_M$) persists, unaffected by both the strain and the local magnetic field induced by MFM tip. This indicates that the maze-like domain phase possesses the lowest free energy dominated by exchange energy and magnetic dipolar interactions. Pre-annealing at 323 K enables the formation of an ordered skyrmion lattice with only 0.80% strain. This transformation is facilitated by the delicate interplay of DMI, magnetic anisotropy, dipolar interactions, and Zeeman effects, effectively lowering the energy barriers and enable the small magnetic field of the tip to nucleate skyrmion lattice.

On the other hand, $Fe^{int}$ atoms also reduce the lattice symmetry and enhances the DMI of FGaT, thus stabilizing the skyrmion lattice[26-27]. At the same time, it should be noted that in high temperature (333 K and 343 K) pre-annealed sample, $Fe^{int}$ atoms introduced local random magnetic interactions within the material. As the temperature decreases from $T_C$, these random interactions drive the formation of stripe dislocations, skyrmions, skyrmionium, and a distinct mixed phase of ferromagnetic and r-ferromagnetic domains. In the r-ferromagnetic phase, the reduced magnetic dipole-dipole interactions allow domain wall energy to dominate, thereby stabilizing the stripe domain structure[26] (Figure S19), thereby restricting the formation of skyrmions by application of strain.

These findings provide significant insights into the mechanical control of magnetic phases in FGaT and underline the critical role of external conditions such as strain and magnetic field in determining the stability and transition dynamics of magnetic textures.

**2.5. Strain-Induced Robust and Field-free Skyrmion Lattice**

After creating an ordered skyrmion lattice by strain, we further investigated the robustness of skyrmion lattice due to its potential application in high-density data storage devices. A series of comparative experiments to study the robustness of strain induced skyrmion lattice, is shown in **Figure 4** and Figure S20-S23. Strain was initially applied to induce the formation of a skyrmion lattice (Figure S20a), which was subsequently released. Notably, the skyrmion lattice remained stable even in the absence of strain (Figure S20b). Upon applying and releasing of 0.80% strain, the lattice structure showed no obvious changes, further emphasizing the inherent stability and robustness of the strain-induced skyrmion lattice (Figure S20c). The FGaT was subjected to a strong external magnetic field to drive the annihilation of the skyrmion lattice (Figure S21a), resulting in the formation of stripe domains upon field removal (Figure S21b). However, repeated scanning with the MFM tip induced the reformation of the skyrmion lattice (Figure S21c). Figure S22 examines whether the robustness of skyrmion lattice relates to pinning by analyzing differences in defects between the reconstructed (Figure S22d-e) and initial (Figure S22a-c) skyrmion lattice. Despite the skyrmion lattice gradually recovering after multiple MFM tip scans, the defect positions do not fully realign post erasure with a strong magnetic field. This indicates the skyrmion lattice stability does not originate from pinning effects. A similar procedure was performed on the sample after releasing strain, which also led to the re-emergence of the skyrmion lattice (Figure S21d-f). These results indicate that once strain is applied to the FGaT crystal, its effect on the skyrmion lattice is irreversible, underscoring the long-term stability of the strain-induced phase.

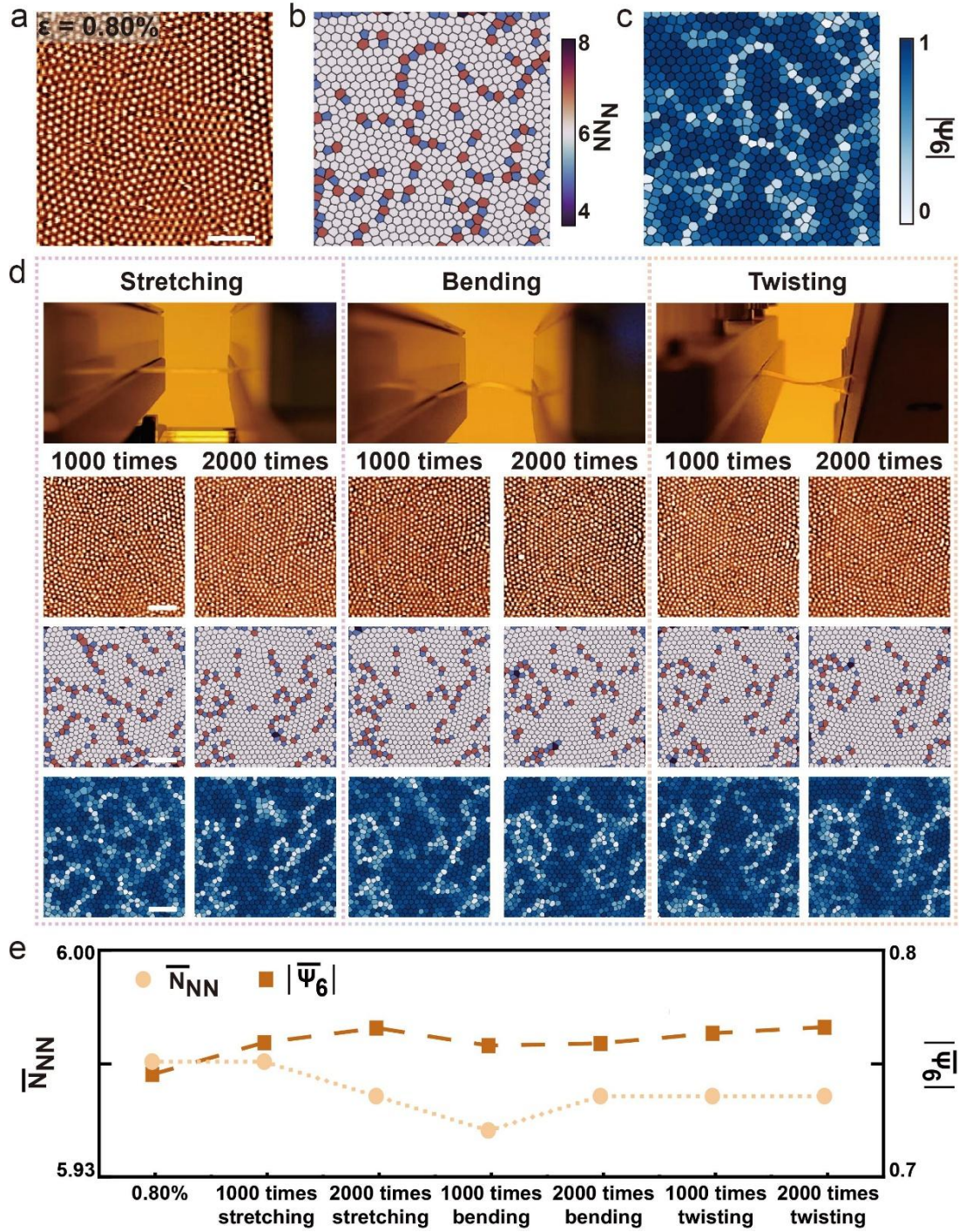

**Figure 4 | Robustness of skyrmion lattice under diverse cyclic strain application. a-c,** Stability of the strain-induced skyrmion lattice under 0.80% strain. The MFM image **(a)**, Voronoi tessellation **(b)** and bond orientational parameter **(c)** of the strain-induced skyrmion lattice. **d,** Series of tests assessing the skyrmion lattice robustness to repeated mechanical deformations: stretching, bending, and twisting, applied 2000 times with steps of 1000 cycles. Voronoi tessellation diagrams, and bond orientational parameter maps are provided for each test condition to evaluate the stability of the skyrmion lattice. **e,** Average Voronoi tessellation $N_{NN}$ and average $|\Psi_6|$ of each state in **d**. Despite extensive application of cyclic strain, minimal changes in skyrmion lattice confirmed by Voronoi tessellation and bond orientational order
17

parameters analysis, demonstrates robustness of skyrmion lattice. Scale bar, 2 $\mu m$.

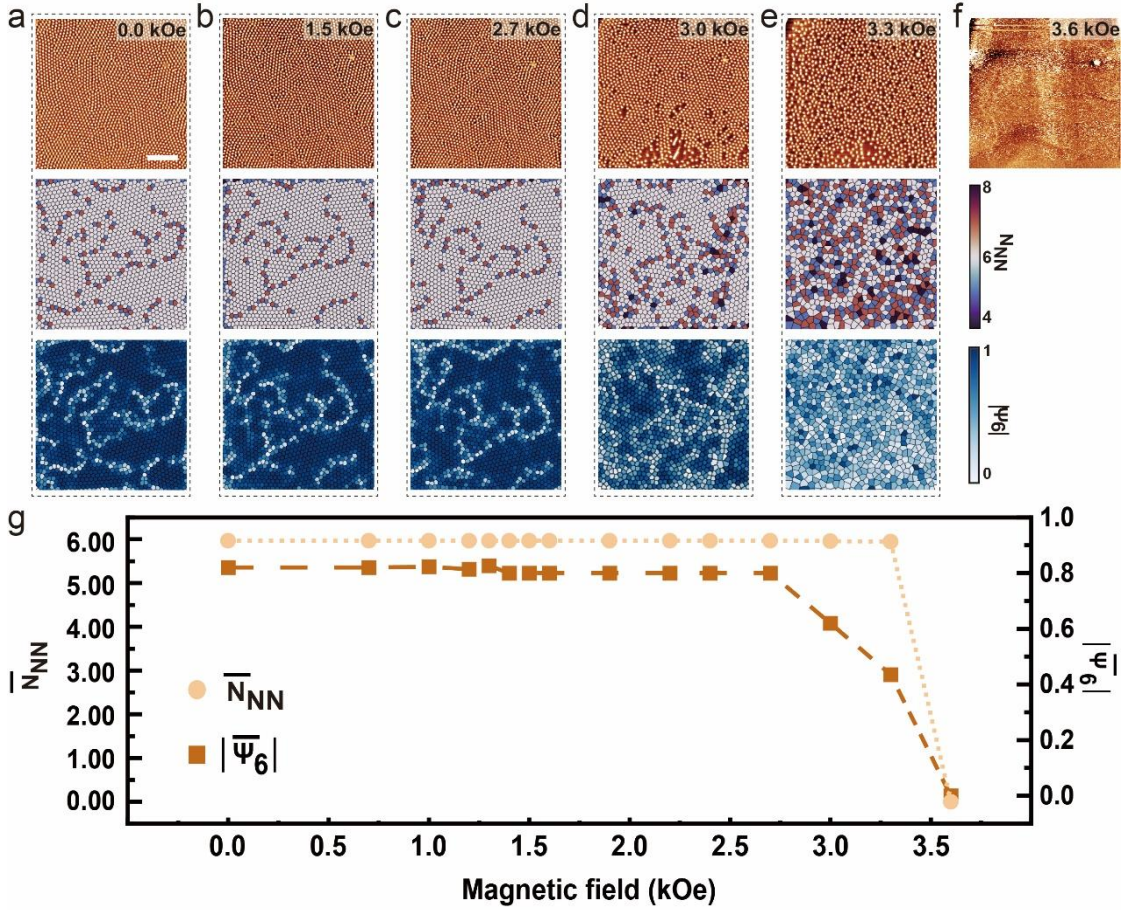

**Figure 5 | Robustness comparison of strain-induced skyrmion lattice under varying external magnetic fields. a–f,** MFM images of the strain-induced skyrmion lattice under increasing magnetic fields, with corresponding Voronoi tessellations and bond orientation parameter analyses. The lattice remains stable up to 2.9 kOe before transitioning into a liquid-like phase at 3.0 kOe and reaching saturation at 3.6 kOe. **g,** Evolution of the average corresponding Voronoi tessellations $N_{NN}$ and average bond orientational order parameter $|\Psi_6|$ as a function of magnetic field for strain-induced skyrmion lattice, confirming the enhanced stability of the strain-induced skyrmion lattice. Scale bar, 2 $\mu m$.

Furthermore, we performed extensive stretching, bending, and twisting tests to assess the robustness and stability of the strain-induced skyrmion lattice, as depicted in Figure 4. Despite undergoing thousands of test cycles, the skyrmion lattice remained stable. To further analyze the skyrmion lattice arrangement, we employed statistical analysis to get Voronoi tessellation diagrams and bond orientation parameter maps. The average Voronoi tessellation $N_{NN}$ and average bond orientational order parameter $|\Psi_6|$ shown in Figure 4d, remains consistently high for the strain-induced skyrmion lattice, indicating enhanced structural integrity. The minor fluctuations in the curve may arise from edge effects in the selected data regions, but the overall variation remains negligible. These analyses confirm that the skyrmion lattice maintains its orderly arrangement even after extensive stretching, bending, and twisting. To further elucidate the stability of the strain-induced skyrmion lattice, measurements under varying external magnetic fields, different external temperature and storage environments were performed, as presented in **Figure 5** and Figure S23-S24. Specifically, the strain-induced skyrmion lattice remains stable up to 2.9 kOe, undergoes a transition into a liquid-like phase at 3.0 kOe, and reaches saturation at 3.6 kOe. Compared with the defective areas, the perfect skyrmion lattice regions can stably exist at higher external temperatures (~323 K). Additionally, the skyrmion lattice demonstrates remarkable temporal stability, maintaining its integrity for extended durations in both inert glove box environments and ambient atmospheric conditions, which further underscores its robustness (Figure S24). The unexpectedly high stability of strain-generated skyrmions makes it possible to design a flexible magnetic field detector (Figure S25). Since the skyrmion lattice state and other magnetic states have different resistances[33], changes in the external magnetic field can be detected by monitoring the change in resistance of device. In addition, this approach remains effective even when the direction of the applied vector magnetic field is altered. It may indicate the potential of FGaT in low-power spintronic devices, particularly in flexible electronics applications.

## 3. Conclusion

In summary, we have demonstrated the potential of uniaxial inhomogeneous strain to precisely manipulate and stabilize ordered skyrmion lattice in 2D vdW FGaT at room temperature, introducing a novel approach for low-power flexible spintronics. These findings were directly imaged using in-situ strain MFM, where skyrmion lattice remained robust even after the strain was removed. The synergy between pre-annealing and applied strain was critical, as it not only facilitated the initial skyrmion lattice formation but also enabled its robustness, which was not achievable by either pre-annealing or magnetic field application alone. In addition, through STEM structural analysis combined with magnetic performance testing analysis, we found that low-temperature pre-annealing led to the precipitation of FeTe$_2$ on the surface of the sample, which disrupted the central symmetry of FGaT, enhanced DMI, and changed other magnetic parameters. The inhomogeneous strain formed by stretching the flexible substrate increases the DMI of the material system and reduces the energy barrier, thereby promoting the formation of skyrmion lattice under local magnetic field of the MFM tip. Our findings may indicate the potential of FGaT in advanced low-power spintronics devices, particularly in flexible spintronics applications.

## 4. Methods

*Fe$_3$GaTe$_2$ single crystal growth*

Fe$_3$GaTe$_2$ crystals were grown using the self-flux method. Highly pure Fe (99.9 %), Ga (99.99 %), and Te (99.99 %) powders, with a mole ratio of 1:1:2, were mixed under Ar atmosphere in a glove box and sealed in a quartz tube under 1.9 Pa. To capture the excess flux during centrifugation, a crucible filled with quartz wool was placed on top of the growth crucible. The melt was homogenized at 1000 °C for approximately 24 hours, then cooled quickly to 880 °C in 1 hour, followed by slow cooling to 780 °C in 100 hours. At this temperature, the ampoules were removed from the furnace and placed in a centrifuge to expel the excess flux. Then large single crystals could be obtained. FGaT nanoflakes were mechanically exfoliated onto a freshly cleaned SiO$_2$/Si substrate.

*Device fabrication and magneto-transport measurements*

A standard Hall bar electrode of 15 nm Cr/ 50 nm Au was pre-fabricated on a 300 nm oxidized layer SiO$_2$/Si substrate by using ultra violet lithography (TuoTuo Technology UV Litho-ACA) and thermal evaporation deposition (VAC COAT DESK THERMAL EVAPORATOR). Then, mechanically exfoliated FGaT nanosheets were transferred onto the Hall bar pattern by using the polydimethylsiloxane stamp.

The electrical transport characteristics are measured using a source meter 2450 (KEITHLEY) at room temperature with four-terminal configuration using the gold wire connected by the bonding machine (WEST·BOND) and utilizing external magnetic field from conventional available attocube system (attoDRY2100) which can add an adjustable magnetic field perpendicular to the sample plane.

For resistance testing, standard electrodes of 15 nm Cr/ 50 nm Au were pre-fabricated on flexible PET substrate by using ultra violet lithography (TuoTuo Technology UV Litho-ACA) and thermal evaporation deposition (VAC COAT DESK THERMAL EVAPORATOR). Then, mechanically exfoliated FGaT nanosheets were transferred onto the electrode by using the polydimethylsiloxane stamp. A Keysight B1500A semiconductor parameter analyzer was used to apply voltage across the devices. The samples were tested in a closed-cycle magneto-optical exchange gas cryostat (attoDRY2100), which was equipped with a superconducting magnet capable of providing 9 T out-of-plane and 3 T in-plane magnetic fields.

*Reflective magnetic circular dichroism (RMCD) measurements*

RMCD measurements were performed in a closed-cycle helium cryostat (attoDRY2100) with a temperature range from 1.8 K to 300 K, a maximum out-of-plane magnetic field of up to 9 T, and an in-plane magnetic field of up to 3 T. The filtered excitation was first passed through an optical chopper at 750 Hz, and then through a 45º linear polarizer and photo elastic modulator (PEM) with a maximum retardance of λ/4. The PEM imparts a sinusoidal phase modulation at 50.06 kHz, allowing for excitation of the sample with alternating left and right circularly polarized light at this frequency. The signal reflected back from the sample was separated from the incidence path using a nonpolarizing beam splitter and sent into photodiode. The photodiode signals of the reflected intensity and the RMCD intensities was fed into lock-in amplifier and analyzed by the Lock-in amplifier.

*Raman Spectroscopy measurements*

The Raman measurements were obtained by a Witec Alpha 300R confocal Raman microscope. The 532 nm linear polarized laser was focused perpendicular to the surface of the sample through a 100x objective lens (numerical aperture = 0.9). The maximum power is kept below 0.2 mW to avoid the laser thermal effects on the sample. The Raman signals were first collected by a photonic crystal fiber and then coupled into the spectrometer with 1800 g/mm grating.

*Magnetic force microscopy measurements (MFM)*

MFM measurements were conducted with the DIMENSION ICON (Bruker). The magnetic field was generated by a neodymium magnet. The measurement was conducted in a glove box protected by high-purity nitrogen gas at room temperature. The magnetic tip from Nanoworld coated with Co/Cr was used for all measurements. The force constant and resonance frequency of this tip are 3 N/m and 75 kHz, respectively. During the measurement, the topography of the sample was first acquired in tapping mode. Then, the MFM tip was lifted by 100 nm above the sample to measure the magnetic signal during the second scan.

*Macroscopic magnetization measurements*

The macroscopic magnetization measurements were carried out using the superconducting quantum interference device MPMS3 magnetometer (Quantum Design).

*High-angle annular dark-field scanning transmission electron microscopy (HAADF-STEM)*

The lamella samples were fabricated from bulk crystals and thinned by the FIB milling technique on an FEI Helios G4 UX. The HAADF-STEM images were obtained using an FEI Titan Cs Probe.

**Supporting Information**

Supporting Information is available from the Wiley Online Library or from the author.

**Acknowledgements**

This work was supported by the National Key Research and Development Program of China (No.2021YFA 1200800) and the Start-up Funds of Wuhan University. We thank Dr. HuiYing Sun from the Core Facility of Wuhan University for the measurement in MPMS. This work was also supported from the Singapore Ministry of Education Tier 3 Programme 'Geometrical Quantum Materials' AcRF Tier 3 (MOE2018-T3-1-002). This research is also supported by the Ministry of Education, Singapore, under its Research Centre of Excellence award to the Institute for Functional Intelligent Materials (Project No. EDUNC-33-18-279-V12).

References

[1] a)S. S. P. Parkin, M. Hayashi, L. Thomas, *Science* **2008**, 320, 190; b)T. Nagai, H. Yamada, M. Konoto, T. Arima, M. Kawasaki, K. Kimoto, Y. Matsui, Y. Tokura, *Physical Review B* **2008**, 78.

[2] a)A. Wachowiak, J. Wiebe, M. Bode, O. Pietzsch, M. Morgenstern, R. Wiesendanger, *Science* **2002**, 298, 577; b)I. A. Malik, H. Huang, Y. Wang, X. Wang, C. Xiao, Y. Sun, R. Ullah, Y. Zhang, J. Wang, M. A. Malik, I. Ahmed, C. Xiong, S. Finizio, M. Kläui, P. Gao, J. Wang, J. Zhang, *Science Bulletin* **2020**, 65, 201.

[3] U. K. Rößler, A. N. Bogdanov, C. Pfleiderer, *Nature* **2006**, 442, 797.

[4] a)A. Fert, V. Cros, J. Sampaio, *Nature Nanotechnology* **2013**, 8, 152; b)A. Fert, N. Reyren, V. Cros, *Nature Reviews Materials* **2017**, 2.

[5] a)J. Jiang, J. Tang, Y. Wu, Q. Zhang, Y. Wang, J. Li, Y. Xiong, L. Kong, S. Wang, M. Tian, H. Du, *Advanced Functional Materials* **2023**, 33; b)Z. Hou, Q. Wang, Q. Zhang, S. Zhang, C. Zhang, G. Zhou, X. Gao, G. Zhao, X. Zhang, W. Wang, J. Liu, *Advanced Science* **2023**, 10; c)T.-E. Park, L. Peng, J. Liang, A. Hallal, F. S. Yasin, X. Zhang, K. M. Song, S. J. Kim, K. Kim, M. Weigand, G. Schütz, S. Finizio, J. Raabe, K. Garcia, J. Xia, Y. Zhou, M. Ezawa, X. Liu, J. Chang, H. C. Koo, Y. D. Kim, M. Chshiev, A. Fert, H. Yang, X. Yu, S. Woo, *Physical Review B* **2021**, 103.

[6] S. Zhang, J. Zhang, Y. Wen, E. M. Chudnovsky, X. Zhang, *Communications Physics* **2018**, 1.

[7] a)B. He, R. Tomasello, X. Luo, R. Zhang, Z. Nie, M. Carpentieri, X. Han, G. Finocchio, G. Yu, *Nano Letters* **2023**, 23, 9482; b)M. Finazzi, M. Savoini, A. R. Khorsand, A. Tsukamoto, A. Itoh, L. Duò, A. Kirilyuk, T. Rasing, M. Ezawa, *Physical Review Letters* **2013**, 110; c)Z. Li, H. Zhang, G. Li, J. Guo, Q. Wang, Y. Deng, Y. Hu, X. Hu, C. Liu, M. Qin, X. Shen, R. Yu, X. Gao, Z. Liao, J. Liu, Z. Hou, Y. Zhu, X. Fu, *Nature Communications* **2024**, 15.

[8] a)S.-I. Park, Y. Xiong, R.-H. Kim, P. Elvikis, M. Meitl, D.-H. Kim, J. Wu, J. Yoon, C.-J. Yu, Z. Liu, Y. Huang, K.-c. Hwang, P. Ferreira, X. Li, K. Choquette, J. A. Rogers, *Science* **2009**, 325, 977; b)T. Someya, Z. Bao, G. G. Malliaras, *Nature* **2016**, 540, 379.

[9] C. Feng, F. Meng, Y. Wang, J. Jiang, N. Mehmood, Y. Cao, X. Lv, F. Yang, L. Wang, Y. Zhao, S. Xie, Z. Hou, W. Mi, Y. Peng, K. Wang, X. Gao, G. Yu, J. Liu, *Advanced Functional Materials* **2021**, 31.

[10] a)K. Shibata, J. Iwasaki, N. Kanazawa, S. Aizawa, T. Tanigaki, M. Shirai, T. Nakajima, M. Kubota, M. Kawasaki, H. S. Park, D. Shindo, N. Nagaosa, Y. Tokura, *Nature Nanotechnology* **2015**, 10, 589; b)T. Koretsune, N. Nagaosa, R. Arita, *Scientific Reports* **2015**, 5; c)Y. Nii, T. Nakajima, A. Kikkawa, Y. Yamasaki, K. Ohishi, J. Suzuki, Y. Taguchi, T. Arima, Y. Tokura, Y. Iwasa, *Nature Communications* **2015**, 6.

[11] D. A. Kitchaev, I. J. Beyerlein, A. Van der Ven, *Physical Review B* **2018**, 98.

[12] G. H. Ahn, M. Amani, H. Rasool, D.-H. Lien, J. P. Mastandrea, J. W. Ager Iii, M. Dubey, D. C. Chrzan, A. M. Minor, A. Javey, *Nature Communications* **2017**, 8.

[13] Z. Dai, L. Liu, Z. Zhang, *Advanced Materials* **2019**, 31.

[14] K. Huang, D.-F. Shao, E. Y. Tsymbal, *Nano Letters* **2022**, 22, 3349.

[15] a)T. Yokouchi, S. Sugimoto, B. Rana, S. Seki, N. Ogawa, S. Kasai, Y. Otani, *Nature Nanotechnology* **2020**, 15, 361; b)S. Bandyopadhyay, J. Atulasimha, A. Barman, *Applied Physics Reviews* **2021**, 8; c)R. Chen, C. Chen, L. Han, P. Liu, R. Su, W. Zhu, Y. Zhou, F. Pan, C. Song, *Nature Communications* **2023**, 14; d)Y. Yang, L. Zhao, D. Yi, T. Xu, Y. Chai, C. Zhang, D. Jiang, Y. Ji, D. Hou, W. Jiang, J. Tang, P. Yu, H. Wu, T. Nan, *Nature Communications* **2024**, 15.

[16] N. S. Gusev, A. V. Sadovnikov, S. A. Nikitov, M. V. Sapozhnikov, O. G. Udalov, *Physical Review Letters* **2020**, 124.

[17] a)P. Huang, T. Schönenberger, M. Cantoni, L. Heinen, A. Magrez, A. Rosch, F. Carbone, H. M. Rønnow, *Nature Nanotechnology* **2020**, 15, 761; b)P. Meisenheimer, H. Zhang, D. Raftrey, X. Chen, Y.-T. Shao, Y.-T. Chan, R. Yalisove, R. Chen, J. Yao, M. C. Scott, W. Wu, D. A. Muller, P. Fischer, R. J. Birgeneau, R. Ramesh, *Nature Communications* **2023**, 14; c)J. Zázvorka, F. Dittrich, Y. Ge, N. Kerber, K. Raab, T. Winkler, K. Litzius, M. Veis, P. Virnau, M. Kläui, *Advanced Functional Materials* **2020**, 30; d)A. R. C. McCray, Y. Li, R. Basnet, K. Pandey, J. Hu, D. P. Phelan, X. Ma, A. K. Petford-Long, C. Phatak, *Nano Letters* **2022**, 22, 7804.

[18] a)D. R. Nelson, B. I. Halperin, *Physical Review B* **1979**, 19, 2457; b)E. P. Bernard, W. Krauth, *Physical Review Letters* **2011**, 107.

[19] C. Xu, X. Li, P. Chen, Y. Zhang, H. Xiang, L. Bellaiche, *Advanced Materials* **2022**, 34.

[20] a)A. O. Leonov, M. Mostovoy, *Nature Communications* **2015**, 6; b)X. Zhang, J. Xia, Y. Zhou, X. Liu, H. Zhang, M. Ezawa, *Nature Communications* **2017**, 8.

[21] a)F. Zheng, F. N. Rybakov, N. S. Kiselev, D. Song, A. Kovács, H. Du, S. Blügel, R. E. Dunin-Borkowski, *Nature Communications* **2021**, 12; b)D. Wolf, S. Schneider, U. K. Rößler, A. Kovács, M. Schmidt, R. E. Dunin-Borkowski, B. Büchner, B. Rellinghaus, A. Lubk, *Nature Nanotechnology* **2021**, 17, 250; c)S. Seki, M. Suzuki, M. Ishibashi, R. Takagi, N. D. Khanh, Y. Shiota, K. Shibata, W. Koshibae, Y. Tokura, T. Ono, *Nature Materials* **2021**, 21, 181.

[22] Y. Wang, J. Wang, T. Kitamura, H. Hirakata, T. Shimada, *Physical Review Applied* **2022**, 18.

[23] a)A. Soumyanarayanan, M. Raju, A. L. Gonzalez Oyarce, A. K. C. Tan, M.-Y. Im, A. P. Petrović, P. Ho, K. H. Khoo, M. Tran, C. K. Gan, F. Ernult, C. Panagopoulos, *Nature Materials* **2017**, 16, 898; b)A. Neubauer, C. Pfleiderer, B. Binz, A. Rosch, R. Ritz, P. G. Niklowitz, P. Böni, *Physical Review Letters* **2009**, 102.

[24] Z. H. Ni, T. Yu, Y. H. Lu, Y. Y. Wang, Y. P. Feng, Z. X. Shen, *ACS Nano* **2009**, 3, 483.

[25] G. Zhang, F. Guo, H. Wu, X. Wen, L. Yang, W. Jin, W. Zhang, H. Chang, *Nature Communications* **2022**, 13.

[26] H. Zhang, Y.-T. Shao, X. Chen, B. Zhang, T. Wang, F. Meng, K. Xu, P. Meisenheimer, X. Chen, X. Huang, P. Behera, S. Husain, T. Zhu, H. Pan, Y. Jia, N. Settineri, N. Giles-Donovan, Z. He, A. Scholl, A. N'Diaye, P. Shafer, A. Raja, C. Xu, L. W. Martin, M. F. Crommie, J. Yao, Z. Qiu, A. Majumdar, L. Bellaiche, D. A. Muller, R. J. Birgeneau, R. Ramesh, *Nature Communications* **2024**, 15.

[27] Y. Wu, Y. Hu, C. Wang, X. Zhou, X. Hou, W. Xia, Y. Zhang, J. Wang, Y. Ding, J. He, P. Dong, S. Bao, J. Wen, Y. Guo, K. Watanabe, T. Taniguchi, W. Ji, Z. J. Wang, J. Li, *Advanced Materials* **2023**, 35.

[28] H. Shi, J. Zhang, Y. Xi, H. Li, J. Chen, I. Ahmed, Z. Ma, N. Cheng, X. Zhou, H. Jin, X. Zhou, J. Liu, Y. Sun, J. Wang, J. Li, T. Yu, W. Hao, S. Zhang, Y. Du, *Nano Letters* **2024**, 24, 11246.

[29] a)C. Zhang, Z. Jiang, J. Jiang, W. He, J. Zhang, F. Hu, S. Zhao, D. Yang, Y. Liu, Y. Peng, H. Yang, H. Yang, *Nature Communications* **2024**, 15; b)S. Jin, Z. Wang, S. Dong, Y. Wang, K. Han, G. Wang, Z. Deng, X. Jiang, Y. Zhang, H. Huang, J. Hong, X. Wang, T. Xia, S.-W. Cheong, X. Wang, *Journal of Materiomics* **2025**, 11.

[30] A. Chakraborty, A. K. Srivastava, A. K. Sharma, A. K. Gopi, K. Mohseni, A. Ernst, H. Deniz, B. K. Hazra, S. Das, P. Sessi, I. Kostanovskiy, T. Ma, H. L. Meyerheim, S. S. P. Parkin, *Advanced Materials* **2022**, 34.

[31] a)S. Xia, etc. https://doi.org/10.48550/arXiv.2505.06924, 2025; b)G. Li, S. Ma, Z. Li, Y. Zhang, Y. Cao, Y. Huang, Advanced Functional Materials 2023, 33; c)Z. Zhang, M. Cai, R. Li, F. Meng, Q. Zhang, L. Gu, Z. Ye, G. Xu, Y.-S. Fu, W. Zhang, Physical Review Materials 2020, 4.

[32] Y. Zhang, J. Liu, Y. Dong, S. Wu, J. Zhang, J. Wang, J. Lu, A. Rückriegel, H. Wang, R. Duine, H.

Yu, Z. Luo, K. Shen, J. Zhang, *Physical Review Letters* **2021**, 127.

[33] a)X. Zhang, Y. Zhou, K. Mee Song, T.-E. Park, J. Xia, M. Ezawa, X. Liu, W. Zhao, G. Zhao, S. Woo, *Journal of Physics: Condensed Matter* **2020**, 32; b)W. Jiang, G. Chen, K. Liu, J. Zang, S. G. E. te Velthuis, A. Hoffmann, *Physics Reports* **2017**, 704, 1; c)S. Mi, J. Guo, G. Hu, G. Wang, S. Li, Z. Gong, S. Jin, R. Xu, F. Pang, W. Ji, W. Yu, X. Wang, X. Wang, H. Yang, Z. Cheng, *Nano Letters* **2024**.